\documentclass[journal,twocolumn,10pt]{IEEEtran}
\usepackage{multirow}
\usepackage{amsmath}
\usepackage{setspace}
\usepackage{amsthm}
\usepackage{amscd}
\usepackage{hhline}
\usepackage[dvips]{graphicx} % !!!!!!!!!!!!!!!!!!!
\usepackage{amssymb}
\usepackage{xcolor}
\numberwithin{equation}{section}
\definecolor{miti}{RGB}{0, 176,80}
\definecolor{prep}{RGB}{237, 125,49}
\definecolor{resp}{RGB}{91, 155,213}
\definecolor{reco}{RGB}{112, 48,160}
\definecolor{grey}{RGB}{245, 245,245}

\begin{document}

%\tableofcontents
% \listoftables
 %\listoffigures

\title{ 
Counter-Epidemiological Projections of e-Coaching 
\thanks{}
}

\author{{Kenneth Lai}, {Svetlana N. Yanushkevich}, and {Vlad P. Shmerko} 
\thanks{
 K. Lai, S. N. Yanushkevich, and V. P. Shmerko
 are with Biometric Technology Laboratory, Department of Electrical and Software
 Engineering, University of
 Calgary, Canada, Web: http://www.ucalgary.ca/btlab. E-mail:
\{kelai,syanshk,vshmerko\}@ucalgary.ca. 
}
\\
%\Large{\bf DRAFT \#01 \today}
}

%\date{}

\markboth{%IEEE TRANSACTIONS ON INFORMATION FORENSICS AND SECURITY, 2019 
}{ \MakeLowercase{\textit{et al.}}:
.....}

\maketitle

\begin{abstract}
 This paper considers e-coaching at times of pandemic. It utilizes the Emergency Management Cycle (EMC), a core doctrine for managing disasters. The EMC dimensions provide a useful taxonomical view for the development and application of e-coaching systems, emphasizing technological and societal issues. Typical pandemic symptoms such as anxiety, panic, avoidance, and stress, if properly detected, can be mitigated using the e-coaching tactic and strategy. In this work, we focus on a stress monitoring assistant developed upon machine learning techniques. We provide the results of an experimental study of a prototype of such an assistant. Our study leads to the conclusion that stress monitoring shall become a valuable component of e-coaching at all EMC phases.
\end{abstract}


%\tableofcontents

%****SECTION*******SECTION*********SECTION*********SECTION****
\section{Introduction}\label{sec:}
%****SECTION*******SECTION*********SECTION*********SECTION****

\IEEEPARstart{T}{his} study's motivation is an on-going adaptation of e-coaching toward potential pandemics, and the lessons learned from COVID-19 pandemic related to mental state of human community \cite{[Varma-2021],[Yan-2021]}.

As stated in \cite{[Andre-2021]}, e-coaching ``may contribute to a better understanding of people's affective responses to the COVID-19 crisis. \ldots If ethical, legal, and social implications are addressed appropriately, affective computing technologies may bring a real benefit to society by monitoring and improving people's mental health''. Typical pandemic symptoms include anxiety, panic, avoidance, and stress. Recent studies on COVID-19 highlighted its impact on the human mental state. In particular, monitoring of COVID-19 related stress globally in 63 countries has shown that over 70\% of the respondents had greater than moderate levels of stress, with 59\% meeting the criteria for clinically significant anxiety and 39\% reporting moderate depressive symptoms \cite{[Varma-2021]}.
	
Stress detection has traditionally been a part of the affect recognition process \cite{[Schmidt-2019]} which includes the detection of emotional states such as sadness, happiness, surprise. Visual cues such as human facial expression are often used in automated approaches to affect detection. Recent advances to pattern recognition theory and techniques have greatly empowered this domain of affect detection. Stress is one of the affective states and can be detected using both visual appearance and physiological signals. 

In our study, we support the assumption that e-coaching should be integrated into a system developed to manage global disaster events \cite{[Joshi-2019],[WHO-2020]}. This system is known as Emergency Management Cycle (EMC) \cite{[WHO-EI]}. The EMC framework is adapted in our study since it provides a systematic counter-epidemic view of e-coaching; this view has been missed in pre-pandemic time. 

The proposed in this paper framework paves the way to a strategic EMC road mapping for the post-pandemic e-coaching technologies. This framework is based on the three technology-society premises: 

\subsubsection*{First premise}
E-coaching resources must be integrated into the standardized four-phase EMC mechanism \cite{[WHO-EI]}:
	\begin{eqnarray*}
		\overbrace{\text{{E-coaching}}}^{\text{\it EMC Projections}}
		\equiv \left\{
		\begin{tabular}{l}
			{Mitigation} phase (\emph{preventing steps}); \\
			{Preparedness} phase (\emph{precaution steps}); \\
			{\textbf{Response}} phase (\textit{\textbf{immediate actions}});\\
			{Recovery} phase (\emph{return to Status Quo}) \\
		\end{tabular} 
		\right .
	\end{eqnarray*}

Currently, society is in the response phase of EMC. Note that embedding a component, such as e-coaching, into the EMC is a very particular process because the EMC doctrine provides only general principles.

\subsubsection*{Second premise}
When designing the e-coaching component, a risk mitigation mechanism should be integrated into the stress-conditional scenarios which are typical in pandemics \cite{[Varma-2021]}. Formally, we aim at minimizing the risks of e-coaching using potential risk impact and its probability: $\texttt{\small <e-coaching risk} \texttt{\small >} = F(\texttt{\small Stress impact}, \texttt{\small Probability}).$ Specifically, a stress detector as a personal device should continuously learn strongly individual features of stress in order to adjust the e-coaching tactic or/and strategy.

\subsubsection*{Third premise} 
Typically, e-coaching system is viewed as a network of wearable and wireless sensors utilized for stress detection. In our approach, stress level of e-coached user must be determined, which helps adjust the coaching tactic and strategy. Hence, an experiment must be set up in order to determine what kind of sensors are useful for this purpose. The performance measure, or measure of usefulness, include accuracy among others. 

These premises are used in reporting our results. In Section \ref{sec:Abbreviations}, among often used abbreviations, a set of most popular in e-coaching sensors is including (red fonts). We introduce the epidemiological content of the e-coaching in Section \ref{sec:E-coaching-content}. The problem, approach, and contribution are formulated in Section \ref{sec:Problem-approach}. The conceptual part of our work, i.e. the EMC profiling of e-coaching drivers, is reported in Section \ref{sec:EMC-projections}. Section \ref{sec:Proof-of-concept} offers a stress detector over various sources of biomedical data. Conclusions, recommendations, limitations, and future work are given in Section \ref{sec:Summary}.

%****SECTION*******SECTION*********SECTION*********SECTION****
\section{Abbreviations}\label{sec:Abbreviations}
%****SECTION*******SECTION*********SECTION*********SECTION****
	
	\vspace{2mm}
	\begin{small} 
		\begin{tabular}{ll}
			\textcolor{red}{\bf ACC} &-- Accelerator\\
			\textcolor{red}{\bf BVP} &-- Blood Volume Pulse\\
			%\textbf{CI} &-- Computational Intelligence\\
			\textcolor{red}{\bf ECG} &-- Echocardiogram\\
			\textcolor{red}{\bf EDA} &-- Electrodermal activity\\
			\textcolor{red}{\bf EEG} &-- Electroencephalogram\\
			\textbf{EMC} &-- Emergency Management Cycle\\
			\textbf{ESI} &-- Epidemiological Surveillance \& Intelligence \\
			\textcolor{red}{\bf EMG} &-- Electromyogram\\
			%\textbf{GAN} &-- Generative Adversarial Network\\
			%\textbf{GISRS} &-- Global Influenza Surveillance \& Response System\\
				%\textbf{R\&D} &-- Research \& Development \\
				\textbf{Res-TCN} &-- Residual-Temporal Convolution Network \\
			\textcolor{red}{\bf RESP} &-- Respiration\\
				%\textbf{R\&D} &-- Research and Development\\
				%\textbf{RNN} &-- Recurrent Neural Network\\
			\textbf{SMA} &-- Stress Monitoring Assistant\\
			\textbf{TCN} &-- Temporal Convolution Network\\
			\textcolor{red}{\bf TEMP} &-- Temperature\\
			%\textbf{WS} &-- Wake-Sleep rhythm\\
			%\textbf{WSN} &-- Wearable Sensor Network\\
		\end{tabular}
	\end{small}

%****SECTION*******SECTION*********SECTION*********SECTION****
\section{ Pre- and post-pandemic E-coaching}\label{sec:E-coaching-content}
%****SECTION*******SECTION*********SECTION*********SECTION****
This section revisits the concept of e-coaching in three coordinates: 1) technological advances, 2) epidemiological challenges, and 3) psychological challenges. 

%---------------------------------------------------------
\subsection{Advances of pre-pandemic e-coaching}
\begin{figure}[!ht]
\begin{center}
\includegraphics[scale=0.45]{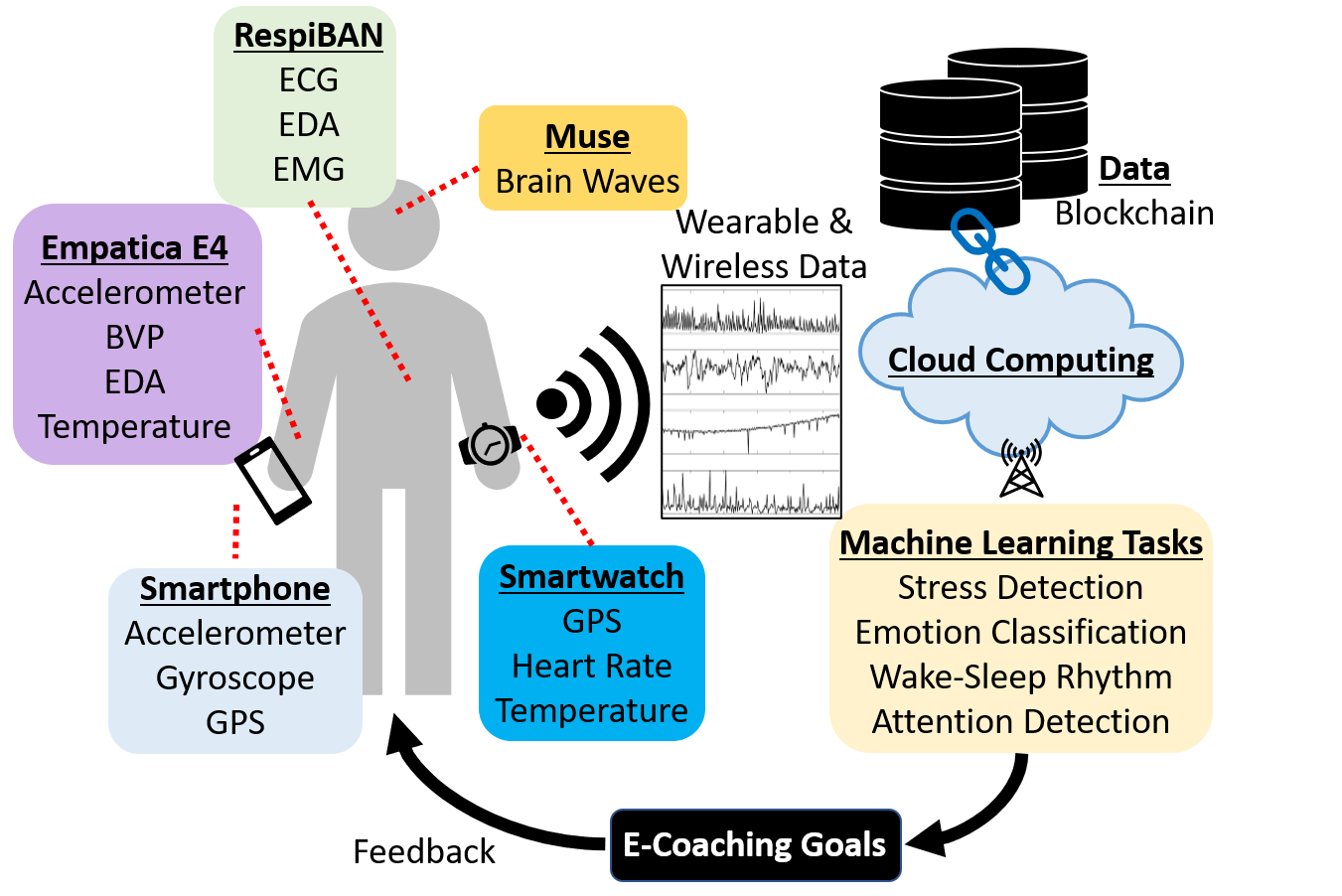}
\end{center}
\caption{The contemporary concept of e-coaching: the perception-action cycle is the core of e-coaching. 1) Wearable sensors sends data to a cloud for processing, 2) Cloud stores transaction of data and processing in a secure blockchain, 3) Cloud performs unique machine-learning tasks such as stress detection, 4) The result of the task is reported and compared to whether it matches the e-coaching goals, and 5) Feedback is provided to the user based on their performance.}\label{fig:cycle}
\end{figure}

%---------------------------------------------------------
Contemporary e-coaching is an integral part of the big technology-society picture such as IoT and smart city \cite{[Celesti-2020]}, as well as cloud computing using blockchain technology \cite{[Melillo-2015]}. Fig. \ref{fig:cycle} illustrates advances in e-coaching. The core of e-coaching is the perception-action cycle aiming at adaptation to the user, e.g. to better self-manage health, health risk, resulting in optimal wellness, improved health outcomes, lowered health risk, and decreased healthcare costs \cite{[National_Society_Health_Coaches-2016]}.

Contemporary e-coaching is defined as an intelligent distributed computing system for performing health coaching tasks, in particular: observe the user, listen, and ask questions, use questioning techniques to identify problems and seeking solutions, facilitate real, lasting positive change, facilitate the exploration of needs, motivations, skills, and thought processes, and support the goal-setting and assessment \cite{[Kamphorst-2017],[Ochoa-2018],van2017coaching,vermetten2020using}. \textbf{At least one of these tasks will be compromised by thestress state of the e-coaching user. }

In e-coaching, the following sources of biomedical information are available using wearable and wireless sensors (Fig. \ref{fig:cycle}):
\begin{itemize}
\item Body sensing networks such as RespiBan\footnote{https://biosignalsplux.com/products/wearables/respiban-pro.html} chest sensor and Empatica E4\footnote{https://www.empatica.com/research/e4/} wrist sensor collect signals like ECG, EDA, EMG, etc.
\item Smartphone sensors such as accelerometer, gyroscope, etc. Applications of smartphones for facial recognition to unlock phones or face analysis for expression recognition.
\item Smartwatch and/or Fitbit\footnote{https://www.fitbit.com/global/en-ca/home} that provides health-related data and analytic such as heart rate and temperature.
\item Muse\footnote{https://choosemuse.com/} headband which reports unique signals such as EEG which can be analyzed for mental activity.
\end{itemize}

An essential trend in e-coaching is the intensive use of deep learning and machine reasoning \cite{[Sannino-2018],[Porumb-2020],[Velikova-2014]}. If new possibilities are opened in e-coaching, this trend leads to the problem of such artificial intelligence bias, interpretability, and explainability of machine reasoning and decision-making \cite{[Gunning-2019],[Whittaker-AI-now-Report-2018]}. 

%---------------------------------------------------------
\subsection{Demand of epidemiological coordinates of e-coaching}
%---------------------------------------------------------
To the best of our knowledge, contemporary e-coaching systems do not take into consideration the variable of epidemiological events. In the reported lessons learned of COVID-19, e.g. \cite{[Alamo-2020],[Costa-2020]}, there no systematic approach for integrating the e-coaching in the Epidemiological Surveillance Intelligence (ESI), and, consequently, into the EMC.

The ESI is a part of any national and global preparedness network such as the Global Influenza Surveillance and Response System (GISRS)~\cite{GISRS}. Epidemiological evidence must be analyzed, monitored, and future scenarios must be predicted. ESI systems have the means to assess a state of epidemic threat in order to introduce suitable countermeasures. In GISRS, for example, the public health decision-making is based on assessing pandemic severity and transmissibility score as guided by Pandemic Severity Assessment Framework. 

Integration of e-coaching into the ESI will be beneficial in many aspects, and will bridge technology-societal gaps in critical epidemic scenarios such as efficient distribution of health care resources (typically limited, as observed in COVID-19 pandemic), delivery of personalized health care (essential as proven in COVID-19 pandemic), and monitoring of health including psychological state (in extreme demand as COVID-19 demonstrated). \textbf{The synergy of e-coaching and ESI results in the creation of controllable counter-epidemic clusters}. Conceptually, the framework of the e-coaching network \cite{[Celesti-2020],[Melillo-2015],[Tseng-2020]} should be comparable with ESI architecture \cite{[Joshi-2019],[WHO-2020]}. 

%---------------------------------------------------------
\subsection{Psychological coordinates of e-coaching}
%---------------------------------------------------------
COVID-19 pandemic has caused an increased level of anxiety, panic, and stress \cite{[Varma-2021],[Yan-2021]}. Psychological states of individuals can be recognized using data from various sources, including visual appearances, facial expressions, voice acoustics, body gestures, as well as body vitals. A demand emerged for automated detection of these emotional states. Counter-epidemic conditions accelerate research in this area, in particular, related to the key source of emotional state, -- facial expression. Pandemic protective measures such as mask-wearing or other coverage significantly decreases the performance of recognizing faces \cite{[Qui-2020]}. This condition has prompted the research of periocular area recognition. In this sub-section, we consider the automated detection of anxiety, panic, and stress using the face and periocular area. These detection approaches are necessary components of post-pandemic e-coaching.

\emph{Avoidance behavior.} 
Anxious individuals show strong avoidance behavior. Avoidance can only be detected using behavioral features and not physiological signals. Hence, a learning mechanism over behavioral bias, e.g. reinforcement learning \cite{[Mkrtchian-2017]}, is needed to detect avoidance: 
\begin{center}
	$\underbrace{\texttt{\small<\emph{Anxiety}>} \longrightarrow
	\texttt{\small<\emph{Avoidance}>}}_{\text{\it E-coaching: Deep Machine Learning}}$
\end{center}

\emph{Anxiety and panic disorder.} 
Monitoring Wake-Sleep (WS) rhythms is an important attribute for e-coaching strategy development and performance evaluation. In \cite{[Jacobson-2021]}, anxiety and panic disorder symptoms are detected and predicted using signals from wake-sleep rhythms captured via wearable sensors and unsupervised deep learning technique (autoencoder):
\begin{center}
	$\underbrace{\texttt{\small<\emph{WS rhythms}>} \longrightarrow
	\texttt{\small<\emph{Psychology disorders}>}}_{\text{\it E-coaching: Deep Machine learning}}$
\end{center} 

\emph{COVID Stress Syndrome.} 
In \cite{[Taylor-2020]}, the notion of COVID Stress Syndrome has been introduced. Authors analyzed a set of factors such as the worry about the dangerousness of COVID-19, worry about the socioeconomic consequences of COVID-19, xenophobic fears, belief in COVID-19-related conspiracy theories, social distancing, poor hand hygiene, and anti-vaccination attitudes. Such a psychological portrait of a pandemic can be useful for stress mitigation.

In our approach, the stress level of the e-coached user is a trigger for changing or adjusting the coaching tactic or/and strategy.

\emph{Boredom.} 
Boredom is one of the most relevant stressors in those who had experienced isolation during the pandemic, e.g. \cite{[Presti-2020]}. Study \cite{[Yan-2021]} provides the following useful insides for developing e-coaching strategies in a pandemic, e.g. links with emotional distress such as:
\begin{eqnarray*}
	\underbrace{\textbf{\texttt{\small Boredom}}
	\Rightarrow \left\{
	\begin{tabular}{l}
		\texttt{\small \emph{Depression, Fear}} \\
		\texttt{\small \emph{Compulsion-anxiety}}\\
		\texttt{\small \emph{Neurasthenia, Hypochondria}} \\
	\end{tabular} 
	\right .}_{\text{\it E-coaching: Deep Machine learning}}
\end{eqnarray*}

\emph{Stress detection/recognition.}
An individual's state of stress heavily influences at least one of the human primary systems, vital for survival: vision, cognitive processing, and motor skill \cite{grossman2008psychological}. Therefore, it is imperative to detect and assess the risk of stress, and, thus, prevent the impact of unwanted effects of e-coaching (measured as probability) using a well-established technique:
$\texttt{\small <Risk>} = F(\texttt{\small Impact},\ \texttt{\small Probability})$, 
where $F$ is a function of the probability of occurrence risk effects and impacts (consequences) \cite{[Lai-2021],[NIST-2017]}.
		
In the EMC projections, stress causes, perception, and manifestation change throughout various EMC phases. This problem addresses a well-identified emerging area of psychology known as \cite{[Engerta-2019],[White-2016]}.
\begin{eqnarray*}
	\underbrace{\textbf{\texttt{\small Stress propagation}}
	\equiv \left\{
	\begin{tabular}{l}
		\texttt{\small \emph{Stress contagion}} \\
		\texttt{\small \emph{Empathic stress}} \\
		\texttt{\small \emph{Stress resonance}}\\
	\end{tabular} 
	\right .}_{\text{\it E-coaching: Deep Machine learning}}
\end{eqnarray*}
Here, stress contagion is defined as catching the emotions of others, and empathic stress is understood as perception of a target individual's state leads to a similar state in an observing individual.

\emph{Stress propagation.}
It is well-understood that stress, fear, and anxiety can be propagated between people. Authors \cite{[Schweda-2021]} study these spread processes in a pandemic. The key conclusion is that it is possible to optimize behavior during the COVID-19 pandemic using various kinds of interventions. Moreover, such interventions could be integrated into the healthcare system during and after the crisis, and foster societal and economic resilience in the long run. In particular, high levels of trait resilience and risk aversion were shown to be most beneficial for the elderly, compared to younger demographics \cite{[McCleskey-2021]}.

\subsection{E-coaching as a cognitive dynamic model}

Inherent properties of the e-coaching fit well into a concept of perception-action cycle explaining the system's adaptation to user \cite{[National_Society_Health_Coaches-2016]}. This perception-action cycle model is reflected, in particular, in a personalized clinical strategy, as well as in continuous health monitoring. This concept is adapted in our approach to modeling the e-coaching system. We proposed that a Stress Monitoring Assistant (SMA) should be integrated into the e-coaching system, using \emph{Haykin's} cognitive dynamic model \cite{haykin2012cognitive}. This model consists of the following components (attributes): \emph{Perception-action cycle} implies that there are perceptor and actuator; \emph{Memory} for the purpose to learn from the environment and store knowledge; \emph{Attention} --  the ability to prioritize the allocation of available resources; and \emph{Intelligence} -- a function that enables the control and decision-making mechanism to help identify intelligent choices. 
 
\begin{figure}[!h]
	\begin{center}
		\includegraphics[width=0.5\textwidth]{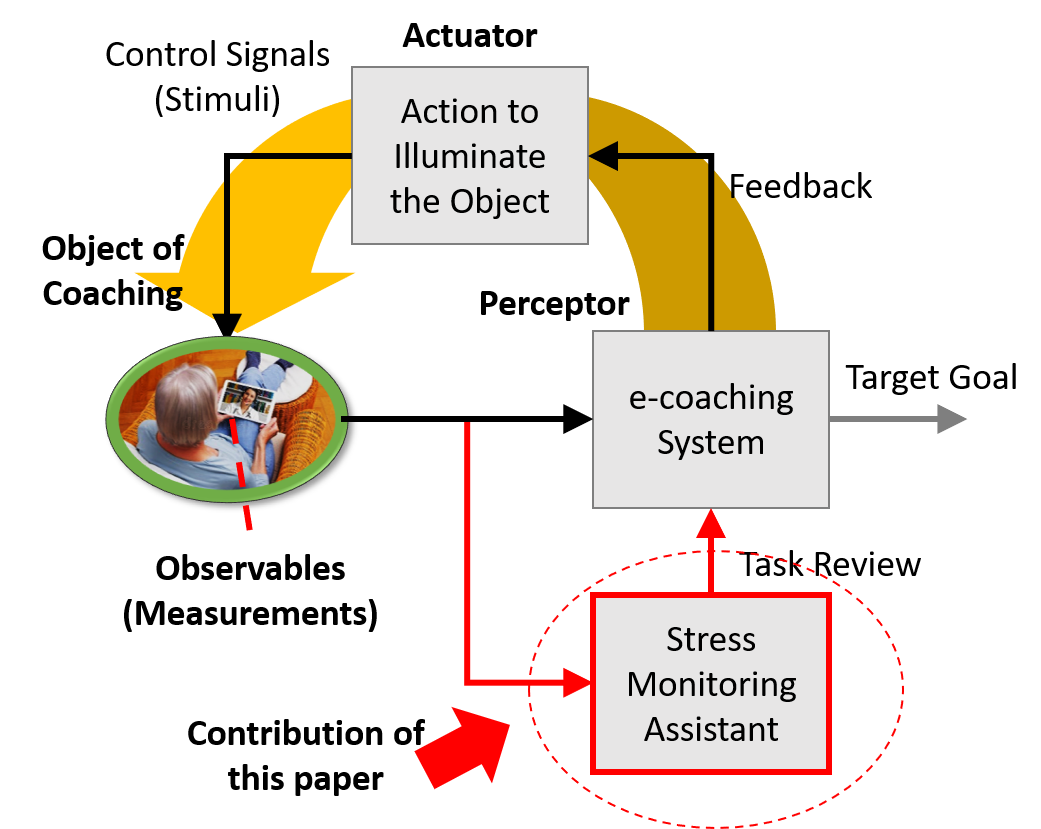}
		\caption{A cognitive dynamic model of e-coaching. Stress level measured by the proposed SMA is a trigger for reviewing the e-coaching tasks and change the tactic and strategy of coaching. }\label{fig:Stress-e-coaching}
	\end{center}
\end{figure}

Fig. \ref{fig:Stress-e-coaching} show how SMA can be integrated into an e-coaching system:
\begin{itemize}
	\item [$-$] The perceived data from an object of coaching are analyzed in order to choose a reasonable strategy and tactic to achieve coaching goals. 
	\item [$-$] The SMA provides emergency indicators for coaching tactics and strategies with respect to the object state.
	\item [$-$] A feedback mechanism measures the difference between a current state and a target state in order to minimize this difference.
\end{itemize}

%****SECTION*******SECTION*********SECTION*********SECTION****
\section{Task formulation and contribution}\label{sec:Problem-approach}
%****SECTION*******SECTION*********SECTION*********SECTION****

\subsection{Task formulation}
The current COVID-19 pandemic is changing the research development priorities in e-coaching. We identify at least two main lessons, or tasks we have learned from the situation:
\begin{enumerate}
	\item []\hspace{-11mm} \textbf{Lesson 1:} E-coaching system design should be adjusted to the realities of the pandemic. The research question is ``How to identify and manage the adjustment tasks?''
	\item []\hspace{-11mm} \textbf{Lesson 2:} Mental health of the e-coached agents (as well as the community in general) is a factor that should be paid particular attention to. The research question is ``How to monitor psychological states of the e-coached agents?''
\end{enumerate}

\subsection{Contributions}
\textbf{Contribution 1:} The e-coaching and counter-epidemic links are determined using the EMC standard. The previous works on e-coaching were limited to a non-epidemic mode. To the best of our knowledge, extreme scenarios of epidemiological attack were out of research interest. Focusing on this gap is the novelty of our approach. 
		
\textbf{Contribution 2:} Psychological (emotional) state of an e-coached agent can be identified using continuous monitoring of personalized psychological and emotional features. This approach is demonstrated by an experimental study of recognition and detection of stress using these features and applying the deep learning networks for their processing and identification. 

Additional value of our approach is a set of recommendations for choosing a type of deep learning tool for e-coaching. 

%****SECTION*******SECTION*********SECTION*********SECTION****
\section{The EMC of e-coaching}\label{sec:EMC-projections}
%****SECTION*******SECTION*********SECTION*********SECTION****

\subsection{The EMC doctrine}
The disaster countermeasure domain consists of all activities on prevention and defense against an epidemiological attack as a rare event. It is reasonable to use a standard and a well-defined taxonomy known as \emph{Emergency Management Cycle} (EMC) \cite{[WHO-EI]}. According to the EMC concept, the e-coaching system should answer the following R\&D questions:
\begin{itemize}
	\item []\hspace{-8mm} How e-coaching can contribute to the mitigation of the disaster? (Mitigation phase);
	\item []\hspace{-8mm} How e-coaching should be integrated into epidemiological surveillance and intelligence? (Preparedness phase);
	\item []\hspace{-8mm} How e-coaching should operate in disaster scenarios? (Response phase); and
	\item []\hspace{-8mm} How e-coaching should reallocate resources in between the modes? ( Recovery phase).
\end{itemize}
 
\begin{figure}[!ht]
	\begin{center}
		\includegraphics[scale=0.5]{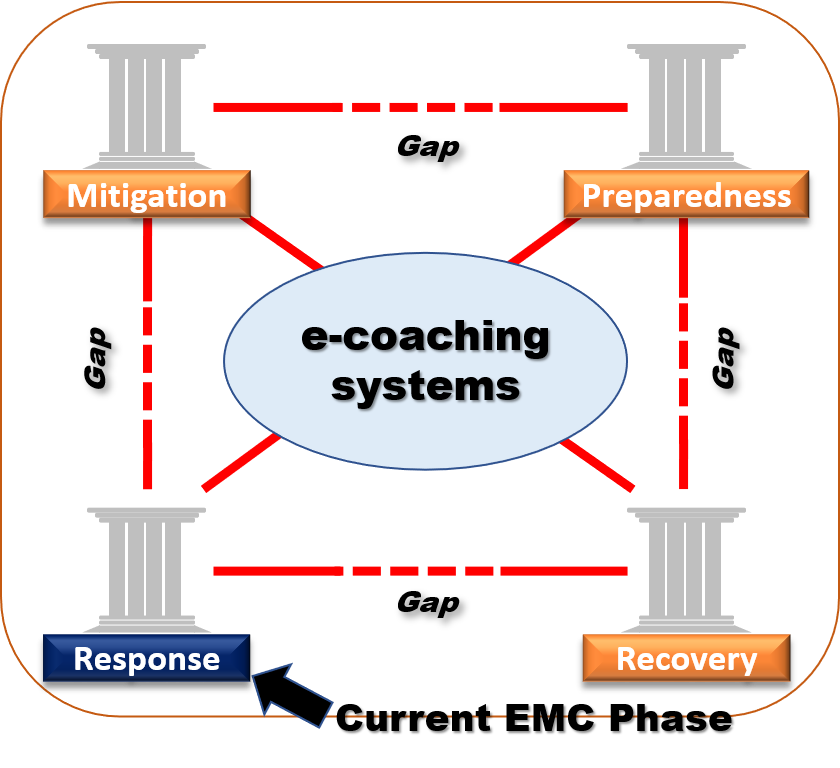}
	\end{center}
	\caption{The EMC of e-coaching: 1) mitigation (preventing steps), 2) preparedness (precaution measures), 3) response (immediate actions), and 4) recovery (actions to return to Status Quo).}
	\label{EMC-gap-e-coaching.png}
\end{figure}

Fig. \ref{EMC-gap-e-coaching.png} provides a taxonomical view of our approach aimed at identifying the technology-societal gaps using the mechanism of the EMC projections of e-coaching. 
	
For example, the \texttt{\emph{Response}} phase we are currently in poses the question: How to return to a normal state (\texttt{\emph{Recovery}} phase) without the losses of achieved efficiency (performance), in preparation for the next potential epidemics (\texttt{\emph{Preparedness}} phase). In this EMC pathway we identified two kinds of technology-societal gaps: 
\begin{eqnarray*}
	\underbrace{\texttt{\emph{Response}}\overset{\text{\it Gap I}}{\longrightarrow}
	\texttt{\emph{Recovery}}\overset{\text{\it Gap II}}{\longrightarrow}
	\texttt{\emph{Preparedness}}
	}_{\text{\footnotesize \it The EMC pathway toward potential epidemics}}
\end{eqnarray*}

 %\subsubsection*{Example 6 (Causality)}
Power of Response is caused by efficiency of Mitigation and Preparedness phases, in particular, how lessons learned from previous epidemics:
\begin{eqnarray*}
	\overbrace{\texttt{\emph{Mitigation}}{\longrightarrow}}^{\text{\footnotesize \it Causal impact}}
	\underbrace{\texttt{\emph{Response}}}_{\text{\footnotesize \it e-coaching}}\overbrace{{\longleftarrow}
	\texttt{\emph{Preparedness}}}^{\text{\footnotesize \it Causal impact}}
\end{eqnarray*}

Alternative approaches to the EMC taxonomical view also exist. For example, a categorization approach based on the COVID-19 pandemic observation was recently proposed in \cite{[Tseng-2020]}. Five categories are distinguished: 1) tracking and predicting virus propagation, 2) characterization of symptoms of virus infections, 3) treatment design, 4) precaution development, and 5) public health policy making. These categories cover all aspects of e-coaching R\&D, e.g. risk profiling and prediction of an individual and groups, integrations of computation intelligence mechanisms and the internet of things (IoT) for smart care, privacy aspects of public health emergencies, and public health policy-making through big data analytics and model simulations.

\subsection{New functions of e-coaching}
According to the EMC doctrine, an e-coaching system should be an integral part of the emergency infrastructure. Hence, the e-coaching system must contribute to the following tasks, in particular:
\begin{itemize}
	\item [$-$]Surveillance and tracking of the infected e-coached users.
	\item [$-$]Predicting virus propagation and pathways.
	\item [$-$]Early detection of infection symptoms.
	\item [$-$]Real-time and early alerting systems for hazardous and forefront outbreaks.
	\item [$-$] Treatment optimization, prediction, and care planning for the best care of the e-coached users.
	\item [$-$]Intelligent analysis of social media and networks for contact tracing and safety control.
\end{itemize}

\subsection{Drivers of e-coaching under EMC profiling }
\begin{figure}[!h]
	\begin{center}
		\includegraphics[width=0.5\textwidth]{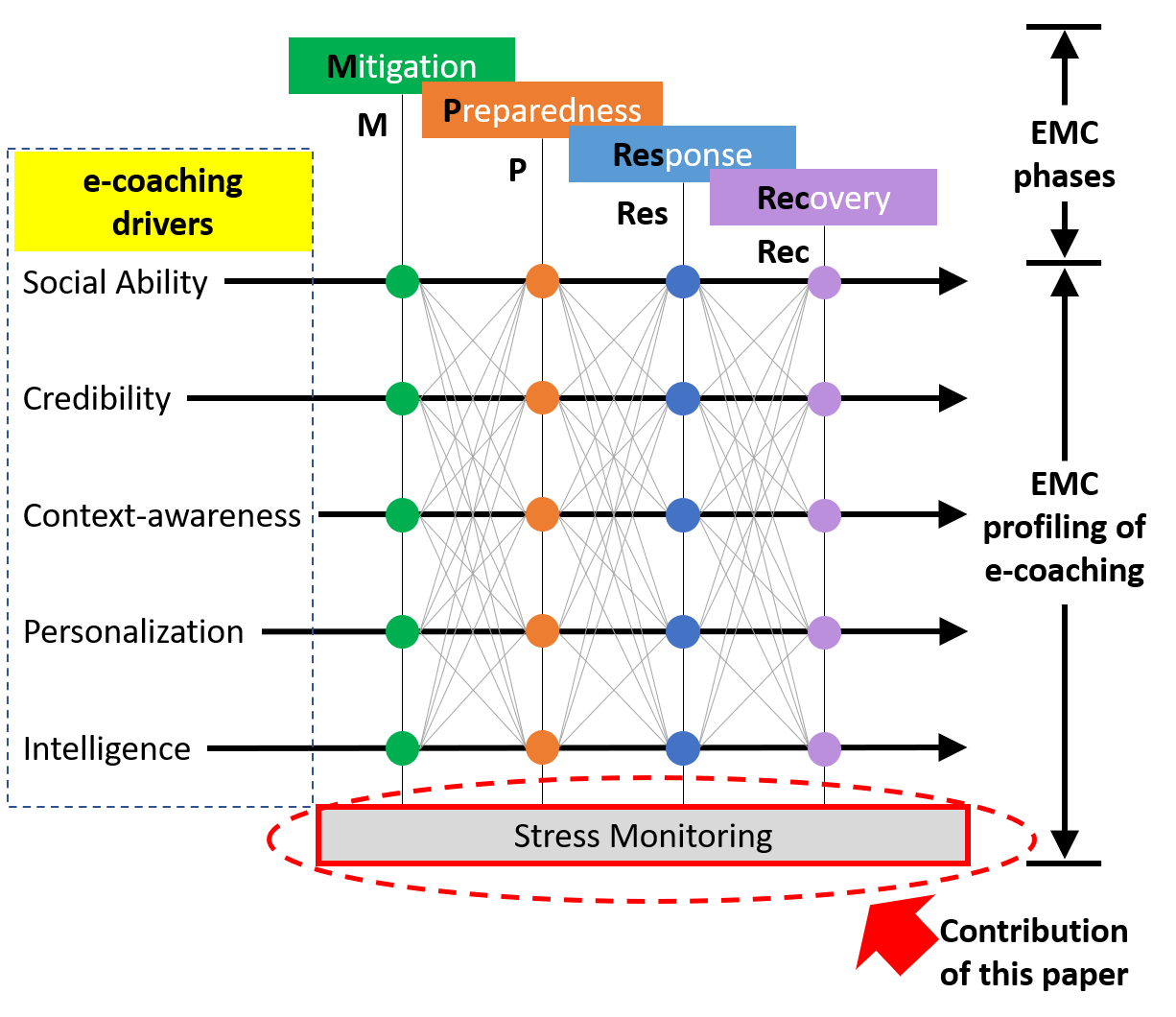}
		\caption{Five e-coaching drivers are differentiated using the four-phase EMC profiling. This $5\times 4$ epidemic-centric e-coaching landscape is monitored by a stress monitoring assistant.}
		\label{fig:EMC-e-coaching}
	\end{center}
\end{figure}

Taxonomical view by \emph{Kamphorst} \cite{[Kamphorst-2017]} has provided feature landscape of e-coaching system grouped in the following categories:
\begin{itemize}
	\item [$-$] \emph{Social ability} in order to establish a collaborative relationship between user and system using conversational technologies.
	\item [$-$] \emph{Credibility and trustworthiness} in interactions between user and system;
	\item [$-$] \emph{Context-aware} in order to stimulate activity, and to assess whether a person's goals are consistent with that person's
	life values;
	\item [$-$] \emph{Personalized} using various feedback mechanisms in order to achieve desired goals;
	\item [$-$] \emph{Intelligence} using machine learning and machine reasoning mechanisms.
\end{itemize}
Our goal is to profile these categories, or e-coaching drivers using the EMC phases. Results are illustrated in Fig. \ref{fig:EMC-e-coaching} as $5\times4$ e-coaching landscape in the EMC projections. Each e-coaching category, or driver is extended accordingly the EMC differentiation as follows:

\subsubsection*{Social ability$^{\textbf{+}}$}
pandemic-centric scenarios of user-machine conversation, e.g. ``How are you feeling after vaccination?'', or Mitigation (M), implementation or Preparedness (P), real-time operation as a part of ESI, or Response (Res), and switch to a regular mode of operation (sometimes called Status Quo), or Recovery (Rec). 
%==============================================================			
\subsubsection*{Credibility and trustworthiness$^{\textbf{+}}$} 
pandemic-centric user-machine interactions, e.g. ``accept additional restriction orders'' (M), support implementation (P), accept real-time operation as a part of ESI (Res), and proceed to recovery (Rec). 
%==============================================================						
\subsubsection*{Context-aware$^{\textbf{+}}$} 
pandemic-centric adjustments user's goals and user's life values (M), contextual measures realization (P), real-time operation at the contextual pace (Res), and localized and timed transitions to Status Quo (Rec). 
%==============================================================						
\subsubsection*{Personalized$^{\textbf{+}}$} 
pandemic-centric adjustments of user responses, e.g. ``Are you sure to make this decision? I detected features of stress.'' (M), user-specific implementation (P), user-adjusted operation as a part of ESI (Res), and individualized restrictions lifting-off (Rec). 
%==============================================================					
\subsubsection*{Intelligence$^{\textbf{+}}$} 
pandemic-centric intelligent support (M), tailored and AI-supported implementation (P), AI-driven real-time operation as a part of ESI (Res), and an AI-assisted switching to Status Quo (Rec). 

This EMC profiling leads to the observation that emerging changes in e-coaching toward integration into the ESI system address the Mitigation (M) and Preparedness (P) phases. Our work is at the frontiers of this process and mainly contributes to the Mitigation phase of the e-coaching stress monitoring.

%--------------------------------------------
\subsection{{Risks of e-coaching}}
%--------------------------------------------
Support individuals in their self-regulation through intelligent e-coaching contains various kinds of risks that should be timely detected. However, mitigation of most risks requires deep intervention in the e-coaching concept. Recent efforts in this direction can be hierarchically taxonomized as follows: 

\subsubsection{Risk of confirmation bias} also known as complacency risk. This is challenging psychological phenomenon of advanced e-coaching theory \cite{[Kamphorst-2020]}. 

\subsubsection{Risk of AI bias} E-coaching support using non explainable and non interpretable AI is risky \cite{[Whittaker-AI-now-Report-2018]}. Machine reasoning using causal networks mostly satisfy these requirements \cite{[Velikova-2014]}.

\subsubsection{ Risk of scenario-centric risk} Contemporary e-coaching systems often use risk monitoring strategy, e.g. \cite{[Tielman-2019],[Khanji-2019]}

\subsubsection{Risk of stress} This risk should be at early warning stage of e-coaching because of user stress state can compromise all efforts on mitigation unwanted coaching effects, -- highlighted by current COVID-19 pandemic; the topic of our study.

Risk recognition detection (recognition among other emotional states) is a vital part of e-coaching. Accordingly NIST \cite{[NIST-2017]}, risk is a ``measure of the extent to which an entity is threatened by a potential circumstance or event, and typically is a function of: (i) the adverse impact, also called cost or magnitude of the harm, that would arise if the circumstance or event occurs, and (ii) the likelihood of event occurrence''. Formally, risk is defined as a function $F$ of impact (or consequences) of a circumstance or event and its occurrence probability, $\texttt{Risk} = F(\texttt{Impact},\ \texttt{Probability})$. If a particular risk factor occurs, then a particular impact on the e-coaching process occurs. The goal of risk management is reducing the probability of the risk factor occurring or/and reducing the impact of this risk factor. 

Consider the e-coaching scenario conditioned by a user stress level. The core of this scenario is a decision on the adjustment of the e-coaching strategy. The trigger for this decision is a result of the stress assessment. This decision is risky because of the possibility of error in the detection of user stress. Risk of acceptance of this decision (which can be correct or incorrect) should be analyzed in terms of its potential impacts or consequences to e-coaching. That is, the risk is a function $F$ of its occurrence probability (i.e. detected user stress) and impact (or consequences):
	
\begin{equation}\label{fig:Risk-adjustment}
	\left\{\hspace{-1.9mm}
	\begin{tabular}{c} 
		\texttt{\small Risk of} \\
		\texttt{\small e-coaching} \\
	\end{tabular} \hspace{-1.9mm}
	\right \}= F(\texttt{\small Impact},\ \underbrace{\texttt{\small Probability}}_{\text{ Stress monitoring}})
\end{equation}

In the described model, \texttt{\small Probability} denotes the probability of successful detection of user stress measured as its \emph{accuracy}, i.e. the number of correctly classified stress patterns among available patterns. In the simplest case, all factors in (\ref{fig:Risk-adjustment}) are rated, for instance, as being low or high.
	
Model (\ref{fig:Risk-adjustment}) provides useful insides for e-coaching with incorporated user's stress detector. The risk of adjustment of the e-coaching strategy is assessed using machine reasoning, similarly to the assessment of the e-coaching risks \cite{[Velikova-2014]}. 	

%****SECTION*******SECTION*********SECTION*********SECTION****
\section{Proof-of-concept experiments}
\label{sec:Proof-of-concept}
%****SECTION*******SECTION*********SECTION*********SECTION****

In our experiment, we adopt the concept of continuous stress monitoring for first responders \cite{[Lai-2021]} for e-coaching users. We argue that in both group of users:
\begin{enumerate}
	\item [$-$]Stress is a personalized psychological state. 
	\item [$-$]An individual's behavior is learned in order to define a personalized stress level, 
	\item [$-$]Deep learning is used as an advanced automated technique for stress detection and recognition, and 
	\item [$-$]Full face analysis is not possible, and only the periocular region is available.
\end{enumerate}

The public-centric and occupational stresses have traditionally been differentiated in terms of their detection, monitoring, and responses.
	 
Occupational stresses have been described in the standards for decision support \cite{dhsCOVID19}. A Wearable Sensor Network (WSN) is a \textbf{preferable} tool for stress detection and monitoring \cite{dias2018wearable} in the workplace and occupational hazard monitoring scenarios. The face is a useful source of data for stress detection \cite{giannakakis2017stress,zhang2020video}; however, in pandemic scenarios, an individual's face may be \textbf{occluded} by personal protective equipment. 

\begin{figure*}[!ht]
	\begin{center}
		\includegraphics[scale=0.5]{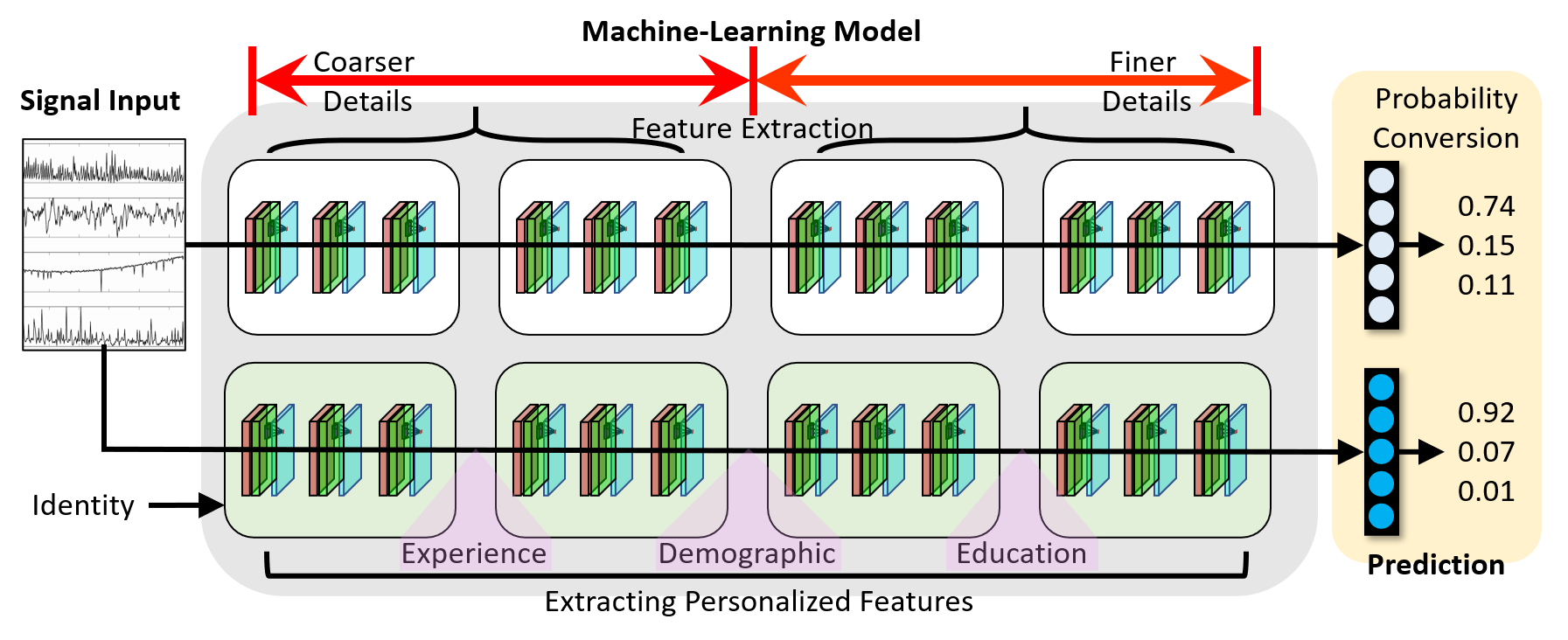}
	\end{center}
	\caption{The core of stress monitoring assistants is the deep learning network of the Res-TCN architecture. The top portion of the network is designed for general purposes while the bottom network is built for personalized processing. Personalized mode unlike the general mode provides the identity information which in some cases shows an increase in performance.}\label{}
\end{figure*}

%--------------------------------------------
\subsection{Experimental scenario}
%--------------------------------------------
A general e-coaching scenario used in the experiments is as follows. Various kinds of physiological signals are available through the WSN, e.g. ECC, EDA, BVP, etc. They can be used for various e-coaching purposes. In our study, we are interested in the detection of stress state as Yes (high level) or No (low level), as well as in recognition of the stress state (among other emotional states). 

%--------------------------------------------
\subsection{Goals of the experiments}
%--------------------------------------------
The primary goal of our experiments is to demonstrate that continuous stress monitoring in e-coaching has been shifted from the category of a 'working idea' to the category of 'prototyping'.

%--------------------------------------------
\subsection{Dataset}
%--------------------------------------------
Among various general requirements to modeling dataset (e.g. synthetic data should be as real as possible, sample sizes must satisfy criteria of statistically representative, and standard protocol in order to repeat experiments), e-coaching emphasizes particular requirements such as multi-sensor data and wearable wireless sensors. Partially satisfies the above requirements is the WESAD dataset, -- Multimodal Dataset for Wearable Stress and Affect Detection by \emph{Schmidt et al.}\cite{schmidt2018introducing}. 

The WESAD dataset was collected from 17 participants, each wearing seven sensors (ACC, ECG, BVP, EDA, EMG, RESP, and TEMP). For each signal different partition is labeled by one of the four different affective states: neutral, stressed, amused, and meditated. There are four different test scenarios: normal, amusement, stress, and meditation. The neutral scenario lasted the first 20 minutes: the participants were asked to do normal activities such as reading a magazine and sitting/standing at a table. In the amusement scenario, the participants watched 11 funny video clips for a total length of 392 seconds. The stress scenario required the participants to perform public speaking and arithmetic tasks for a total of 10 minutes. The last scenario involved a guided meditation session of 7 minutes in duration. The ground truth labels for the affect states were collected using the Positive and Negative Affect Schedule (PANAS) scheme \cite{watson1988development}, upon completion of each trial.

%--------------------------------------------
\subsection{Metrics}
%--------------------------------------------
To assess the classification algorithm performance, it is important to determine the most suitable performance indicators. In the case of balanced data, the traditional metrics include:

\begin{itemize}
	\item []\hspace{-7.5mm}$TP$ -- True Positives (correct predictions of emotion),
	\item []\hspace{-7.5mm}$FN$ -- False Negatives (incorrect predictions of emotion),
	\item []\hspace{-7.5mm}$TN$ -- True Negatives (correct rejections of emotion), and 
	\item []\hspace{-7.5mm}$FP$ -- False Positives (incorrect predictions of emotion). 
\end{itemize}
These numbers form a $2\times 2$ confusion matrix:
\begin{eqnarray*}
	\texttt{Confusion matrix}=
	\bordermatrix{
		& & \cr \noalign{
			\vskip 2pt}
		\mbox{}& TP&FP \cr \noalign{
			\vskip 2pt}
		\mbox{}& FN&TN \cr \noalign{
			\vskip 2pt}
	}
\end{eqnarray*}

These numbers are used to derive accuracy, recall, precision, receiver operating characteristics, and balanced $F_1$-score \cite{luque2019impact}. We selected a few for evaluation, i.e. the accuracy measure and $F_1$-score:
\begin{eqnarray*}
	\texttt{Accuracy}&=&\frac{\text{\textit{TP+TN}}}{\text{\textit{TP+FN+TN+FP}}}\label{eq:acc}\\
	\texttt{\small $F_1$-score}&=&2\times\frac{ \texttt{\small Precision} \times \texttt{\small Recall}}{\texttt{\small Precision} + \texttt{\small Recall}}\label{eq:f1}
\end{eqnarray*}
where \texttt{Recall} (also known as sensitivity) represents the system's ability to detect a specific emotion, $\texttt{Recall}={TP}/{(TP+FN)}$, and \texttt{Precision} (also called positive predictive value) is the system's ability to be correct on a predicted emotion: $\texttt{Precision}={TP}/{(FP+TP)}$.

Accuracy reflects the number of correctly classified patterns among the samples, and, thus, it is a \emph{probability of success} in recognizing the right class of an instance. However, in the case of highly imbalanced datasets, the accuracy measure (\ref{eq:acc}) is \textbf{misleading}. A classifier that is very effective in predicting the majority class, but misses most of the minority classes, may easily have very high accuracy \cite{maratea2014adjusted}.

The $F_1$-score is a weighted average (harmonic mean) of precision and recall rates, representing the system's balanced ability to detect a specific emotion correctly. The $F_1$-score reaches 1 at perfect precision and recall, and 0 at the worst of both \cite{luque2019impact,hand2018note}. This measure provides a way of combining the recall and the precision in order to capture both.

\subsection{Choosing a deep learning network}
Table \ref{tab:e-coaching-dln} provides an overview of the more common issues regarding four main types of deep learning networks. This is one of the first attempts to develop the recommendations for choosing a deep learning mechanism for analyzing emotional states.

The first row represents one of the most common deep learning structures, the Convolutional Neural Network (CNN). A CNN is comprised of two main components, the feature extractor using layers of convolution and the classifier using fully-connected layers of neurons. By default, the main operation of CNNs is convolutions which is excellent at extracting features from images. This task fits well for image-based tasks including face detection, face and face expression recognition.

The second row represents a relatively new deep learning network, the Temporal Convolutional Network (TCN). TCN offers a solution to quickly learn patterns that may exist in time-series data. TCN consists of a series of causal 1D convolution layers optimized for sequential data. The causal nature is well-suited for time-based data processing, including activity recognition, stress detection, and early-warning prediction systems.

The third row represents a well-known technique developed for sequential data, the Recurrent Neural Network (RNN). RNN is composed of a series of cell units that are designed to retain the memory of past events which allows for a more developed prediction. This network is well-suited for language-based processing such as natural language processing, machine translation, and speech recognition.

The last row describes a current technique used for synthesis, the Generative Adversarial Network (GAN). GANs deploy a two-part system consisting of a generator and a discriminator. The generator attempts to generate an image that is indistinguishable from a real image, while the discriminator tries to determine whether the image received is fake or real. Together, these two components play against each other to reach an equilibrium in the process of creating an image that is determined to be real 50\% of the time. This deep learning technique is suited for signal or image synthesis tasks such as generating synthetic voice, face, and conversion of images.

TCN is chosen for the task of emotion classification because of the following reasons:
\begin{itemize}
	\item \emph{Classification task}: Since the goal of this paper is to perform emotion classification and not image generation, the GANs are not suitable.
	\item \emph{Time-series data}: Our input data are physiological signals such as ECG which are time-based data points, the TCN and RNN are best for such data type.
	\item \emph{Time complexity}: Due to the nature of convolution in TCN, the process is consistent and easier to parallelize, as opposed to RNN which requires the previous step to be finished before performing the next operation.
	\item \emph{Memory}: The TCN requires much less memory/parameters compared to RNN when processing the long input sequences. In addition, the TCN can obtain a specific receptive field based on the number of the residual blocks, while the RNN always uses the maximum length of the sequence.
\end{itemize}

\begin{table*}[!htpb]
	\caption{Deep Learning Networks for e-Coaching: Criteria for Choosing.}
	\label{tab:e-coaching-dln}
	\begin{small} 
		\begin{center}
		\begin{tabular}{cccc}
\multicolumn{1}{c}{\normalsize \bf Model}	&		
\multicolumn{1}{c}{\normalsize \bf Challenges}	&
\multicolumn{1}{c}{\normalsize \bf Limitations}	&
\multicolumn{1}{c}{\normalsize \bf Requirements}\\
\multicolumn{4}{c}{\normalsize \colorbox{black}{\color{white}{\bf{\texttt{Convolutional Neural Network}}}}\vspace{.5mm}}\\
\begin{parbox}[h]{0.2\linewidth}{
	\includegraphics[width=0.9\linewidth]{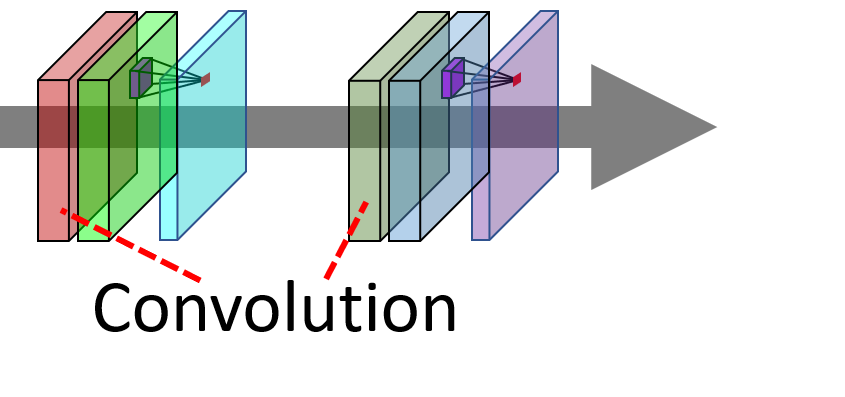}	
}\end{parbox}&	\hspace{-10mm}
\colorbox{grey}{\begin{parbox}[h]{0.24\linewidth} {	
	\item [$\square$] To reduce Over-fitting, apply batch normalization, dropout layers, and L1/L2 regularization.
	\item [$\square$] Under-fitting occurs when using various CNN architectures such as Inception V3, ResNet, and/or HRNet. To reduce under-fitting, use of pre-trained networks or transfer-learning techniques to have better weight initialization.
}\end{parbox}}& \hspace{-1mm}
\colorbox{white}{\begin{parbox}[h]{0.24\linewidth} {	
\begin{itemize}
	\item [$\square$] Choose large enough Sample size, in order to train the network to properly learn the true distribution of the population.
	\item [$\square$] Mitigate possible Bias in data which means that data contains skewed distribution, leading to better representation of the selected classes while minority classes are not well described.
\end{itemize}
}\end{parbox}}& \hspace{-1mm}
\colorbox{grey}{\begin{parbox}[h]{0.24\linewidth} {	
\begin{itemize}
	\item [$\square$] It may work better on Image and/or matrix-based data, since it was historically optimized for such data.
	\item [$\square$] They are appropriate for face recognition \cite{hu2015face}, face expression recognition \cite{li2020attention}, and visual stress detection.
\end{itemize}
}\end{parbox}}\\
\multicolumn{4}{c}{\normalsize \colorbox{black}{\color{white}{\bf{\texttt{Temporal Convolutional Network}}}}\vspace{.5mm}}\\
\begin{parbox}[h]{0.2\linewidth}{
\includegraphics[width=0.9\linewidth]{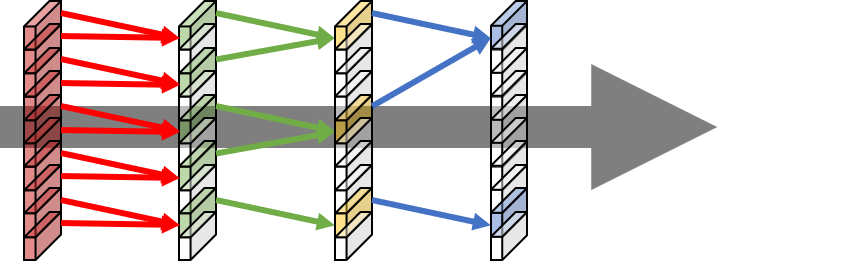}	
}\end{parbox}&	\hspace{-10mm}
\colorbox{white}{\begin{parbox}[h]{0.24\linewidth} {	
\item [$\square$] \emph{Parameter-turning}: Each type of data requires a weight optimized for that specific type of data as each data type behaves differently.
\item [$\square$] \emph{Architecture-search:} Different components are optimized for different tasks.
}\end{parbox}}& \hspace{-1mm}
\colorbox{grey}{\begin{parbox}[h]{0.24\linewidth} {	
\begin{itemize}
	\item [$\square$] \emph{Data quality:} noise in signals deeply impacts the usability of the sample.
	\item [$\square$] \emph{Data variance:} differences for the same type of signal is minimal.
\end{itemize}
}\end{parbox}}& \hspace{-1mm}
\colorbox{white}{\begin{parbox}[h]{0.24\linewidth} {	
\begin{itemize}
	\item [$\square$] It works best for time-series-based data
	\item [$\square$] It is useful for detection of activity using skeletal joint movements \cite{kim2017interpretable}, classification of stress \cite{feng2019dynamic}, and early predictions \cite{catling2020temporal}.
\end{itemize}
}\end{parbox}}\\
\multicolumn{4}{c}{\normalsize \colorbox{black}{\color{white}{\bf{\texttt{Recurrent Neural Network}}}}\vspace{.5mm}}\\
\begin{parbox}[h]{0.2\linewidth}{
\includegraphics[width=0.9\linewidth]{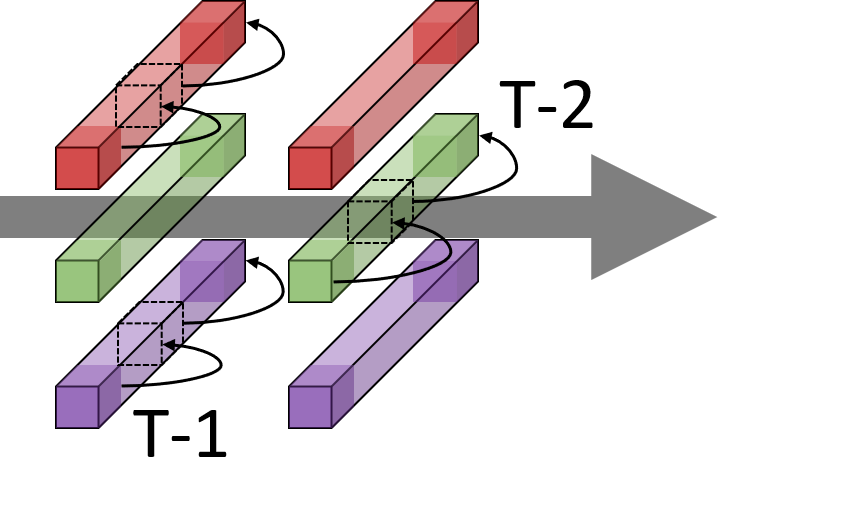}	
}\end{parbox}&	\hspace{-10mm}
\colorbox{grey}{\begin{parbox}[h]{0.24\linewidth} {	
			\item [$\square$] \emph{Vanishing gradient:} Weight updates are diminished every iteration for each recurrent node up to the point where the weights at the bottom layers are no longer updated. Explore different architecture of RNN to reduce the impact of vanishing gradients such as long-short-term-memory cells.
}\end{parbox}}& \hspace{-1mm}
\colorbox{white}{\begin{parbox}[h]{0.24\linewidth} {	
			\begin{itemize}
				\item [$\square$] \emph{Memory:} Cannot remember patterns from a very-long ago.
				\item [$\square$] \emph{Training complexity:} The architectures with recurrent cells is difficult to parallelize.
			\end{itemize}
}\end{parbox}}& \hspace{-1mm}
\colorbox{grey}{\begin{parbox}[h]{0.24\linewidth} {	
			\begin{itemize}
				\item [$\square$] It works best for sequential data.
				\item [$\square$] It was shown effective for natural language processing \cite{mikolov2011extensions}, speech recognition \cite{graves2014towards}, behavior analysis \cite{brattoli2017lstm}.
			\end{itemize}
}\end{parbox}}\\

\multicolumn{4}{c}{\normalsize \colorbox{black}{\color{white}{\bf{\texttt{Generative Adversarial Networks}}}}\vspace{.5mm}}\\
\begin{parbox}[h]{0.2\linewidth}{
\includegraphics[width=0.9\linewidth]{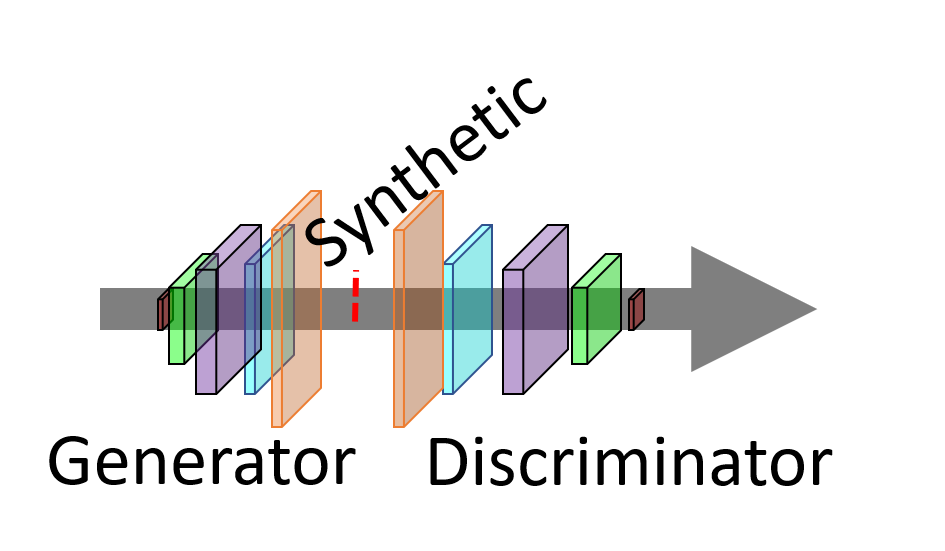}	
}\end{parbox}&	\hspace{-10mm}
\colorbox{white}{\begin{parbox}[h]{0.24\linewidth} {	
			\item [$\square$] \emph{Sample size:} Introduce augmentation techniques to artificially bolster the number of samples for training. Common techniques include rotation, resampling, flipping, and padding.
			\item [$\square$] To avoid under-fitting: Employ different backbone architectures or use of pre-trained networks.
}\end{parbox}}& \hspace{-1mm}
\colorbox{grey}{\begin{parbox}[h]{0.24\linewidth} {	
			\begin{itemize}
				\item [$\square$] \emph{Diversity in data:} Targeted data only contains a specific type of attribute that is maximized by the network, while a more generalizable attribute is desired.
				\item [$\square$] \emph{Training stability:} Extreme sensitivity to the choice of hyper-parameters that can greatly alter the resulting outputs.
			\end{itemize}
}\end{parbox}}& \hspace{-1mm}
\colorbox{white}{\begin{parbox}[h]{0.24\linewidth} {	
			\begin{itemize}
				\item [$\square$] It requires large amount of data in order to capture the true distribution of the targeted sample.
				\item [$\square$] It is effective for data synthesis: it generates synthetic images such as faces \cite{wang2018face}, or synthetic signals such as emotional speech \cite{chang2017learning}, and ECG \cite{golany2019pgans}.
			\end{itemize}
}\end{parbox}}\\
		\end{tabular}
		\end{center}	
	\end{small}
\end{table*}

\subsection{Experiment I: Identification of a stressed user in the e-coached team }
In this paper, we focus on individual stress levels, specifically, how we can improve emotion classification using personalized data. We deploy the analysis of physiological signals in order to detect and identify the level of stress. 

Before analyzing the personalized data, we must determine whether the provided data can be used for subject identification, that is, given a sample of accelerometer data, can we identify the subject? This is a vital task for personalized stress detection as it links subjects to their corresponding stress levels. In Table \ref{tab:id}, we report the accuracy and $F_1$-score of using various physiological signals for subject identification. In this table, the performance is reported for 10 modalities, including 6 signals from the chest region and 4 from the wrist region. The performance measures are calculated using 10-fold cross-validation.

\subsubsection*{Observation 1 (Highest performance)}
The highest performance is obtained via the RESP signal collected from the chest region with an accuracy of 99.84\%. The next highest performing signal is the BVP signal collected from the wrist region with an accuracy of 99.60\%:
\begin{eqnarray*}
	\underbrace{\texttt{\small Maximal Accuracy}}_{\text{\it Stress state signal}}
	= \left\{
	\begin{tabular}{l}
		{\small 99.84	$\pm$	0.10,~\emph{RESP, Chest}} \\
		{\small 99.60	$\pm$	0.19,~ \emph{BVP, Wrist}} \\
	\end{tabular} 
	\right .
\end{eqnarray*}

These results suggest that it is possible to recognize the identity of the sensor wearer given these two types of signals. The worst performing signals are the TEMP signal from both the chest and wrist sensor. Analysis of the best and worst contenders of identification confirms a hypothesis that the more unique the signal becomes, the better for subject identification. Signals that provide common data such as temperature do not offer much information for subject identification.

\begin{table}[!htb]
	\centering
	\caption{E-coaching scenario: Subject Identification Performance using Physiological Signals}
	\label{tab:id}
	\begin{footnotesize}
		\begin{tabular}{@{}cc|cc@{}} 
			\multicolumn{2}{c|} {\small Modality} & \small Accuracy & \small $F_1$-score \\
			\hline
			\hline
			\multirow{6}{*}{\small Chest}	&	ACC	&	88.47	$\pm$	2.05	&	88.51	$\pm$	2.02	\\
			&	ECG	&	97.37	$\pm$	1.44	&	97.38	$\pm$	1.41	\\
			&	EDA	&	60.28	$\pm$	2.45	&	57.09	$\pm$	2.77	\\
			&	EMG	&	20.11	$\pm$	4.30	&	13.68	$\pm$	4.81	\\
			&	RESP	&	99.84	$\pm$	0.10	&	99.84	$\pm$	0.11	\\
			&	TEMP	&	19.15	$\pm$	1.68	&	11.04	$\pm$	2.73	\\
			\hline
			\multirow{4}{*}{\small Wrist}	&	ACC	&	96.92	$\pm$	0.71	&	96.92	$\pm$	0.71	\\
			&	BVP	&	99.60	$\pm$	0.19	&	99.60	$\pm$	0.19	\\
			&	EDA	&	51.08	$\pm$	2.75	&	47.65	$\pm$	3.67	\\
			&	TEMP	&	24.03	$\pm$	3.35	&	18.87	$\pm$	3.66	\\
			\hline
		\end{tabular}
		\end{footnotesize}
\end{table}

\subsection{Experiment II: Stressed e-coaching classification}
Once the identity of the wearer is determined, the next step is to perform the general and personalized emotion classification. The general emotion classification is done in this paper via leave-one-subject-out cross-validation. This form of cross-validation evaluates the performance of the system when one specific subject's data is never seen by the machine-learning model. The purpose of such evaluation is to analyze the system response to unknown data. The personalized emotion classification is then done via 10-fold cross-validation. 10-fold cross-validation measures the performance of the system when each subject's data is partially shown to the machine-learning model. This validation procedure reports the result when the user knows the identity of the subject.

\begin{table}[!htb]
	\centering
	\caption{E-coaching scenario: Emotion Classification Performance using Physiological Signals: (a) Generalized and (b) Personalized}
	\label{tab:emo}
	\begin{footnotesize}
	\begin{tabular}{@{}c|cc|cc@{}} 
		\multicolumn{1}{c|} {\small}& \multicolumn{2}{c|}{\textbf{\small Generalized Mode}} &
		\multicolumn{2}{c}{\textbf{\small Personalized Mode}}\\ 
		\multicolumn{1}{c|} {\small Modality} & \small Accuracy & \small $F_1$-score & \small Accuracy & \small $F_1$-score\\
		\hline
		\hline
		\multicolumn{5}{c}{\small Chest}	\\
		ACC	&	71.2	$\pm$	13.2	&	66.6	$\pm$	15.7	&	84.7	$\pm$	3.8	&	85.2	$\pm$	3.5	\\
		ECG	&	72.7	$\pm$	13.4	&	68.4	$\pm$	16.9	&	92.6	$\pm$	5.7	&	92.6	$\pm$	5.6	\\
		EDA	&	68.6	$\pm$	20.5	&	64.1	$\pm$	24.6	&	60.3	$\pm$	1.8	&	62.0	$\pm$	1.6	\\
		EMG	&	67.9	$\pm$	11.7	&	58.9	$\pm$	14.3	&	56.0	$\pm$	1.4	&	44.9	$\pm$	4.1	\\
		RESP	&	82.9	$\pm$	9.2	&	81.4	$\pm$	9.4	&	99.8	$\pm$	0.1	&	99.8	$\pm$	0.1	\\
		TEMP	&	75.0	$\pm$	10.3	&	67.7	$\pm$	11.6	&	56.8	$\pm$	0.8	&	47.0	$\pm$	2.7	\\
		\hline
		\multicolumn{5}{c}{\small Wrist}	\\
		ACC	&	73.74	$\pm$	16.6	&	72.4	$\pm$	17.4	&	97.3	$\pm$	0.5	&	97.4	$\pm$	0.5	\\
		BVP	&	76.0	$\pm$	10.4	&	72.5	$\pm$	12.6	&	99.6	$\pm$	0.2	&	99.6	$\pm$	0.2	\\
		EDA	&	67.5	$\pm$	18.2	&	63.8	$\pm$	18.9	&	65.5	$\pm$	4.0	&	66.0	$\pm$	4.1	\\
		TEMP	&	59.2	$\pm$	7.1	&	46.3	$\pm$	11.1	&	54.6	$\pm$	1.3	&	42.5	$\pm$	4.6	\\
		\hline
	\end{tabular}
	\end{footnotesize}
\end{table}

Table \ref{tab:emo} reports the emotion classification performance for (a) generalized mode and (b) personalized mode. For each mode, 10 different signals are used for emotion classification, including the accelerometer data, the temperature of the chest, blood volume pulse rate, and electrodermal activity signal measured at the wrist.

\subsubsection*{Observation 2 (Resp- and BVP-centric monitoring)} 
In the general mode, the best performing signal is respiration (Resp) and BVP for the chest and wrist, respectively. 

\subsubsection*{Observation 3 (Comparison)}
An interesting note is that these results coincide with the identification results. The biggest contrast between Table \ref{tab:id} and Table \ref{tab:emo}(a) is that the TEMP signal performs much better at emotion classification than subject identification. In the personalized mode, it is once again the RESP and BVP signals that offer the highest performance with an accuracy of 99.8\% and 99.6\% for the sensors located at the chest and wrist, respectively.

\subsubsection*{Observation 4 (RESP-centric monitoring)} 
For the RESP signal, the general mode is characterized with an accuracy of 82.9\%. This is boosted to 99.8\% if the identity of the wearer is known. 
\begin{eqnarray*}
	\underbrace{\texttt{\small Accuracy \textbf{RESP}}}_{\text{\it Stress state signal}}
	= \left\{
	\begin{tabular}{l}
		{\small 82.9	$\pm$	9.2,~\emph{Generalized Mode}} \\
		{\small 99.8	$\pm$	0.1,~ \emph{Personalized Mode}} \\
	\end{tabular} 
	\right .
\end{eqnarray*}

\subsubsection*{Observation 5 (TEMP-centric monitoring)}
TEMP signal recorded using the chest sensor shows an accuracy of 75.0\% in the general mode, and the accuracy decreases to 56.8\% in the personalized mode. 
\begin{eqnarray*}
	\underbrace{\texttt{\small Accuracy \textbf{TEMP}}}_{\text{\it Stress state signal}}
	= \left\{
	\begin{tabular}{l}
		{\small 75.0	$\pm$	10.3,~\emph{Generalized Mode}} \\
		{\small 56.8	$\pm$	0.8,~ \emph{Personalized Mode}} \\
	\end{tabular} 
	\right .
\end{eqnarray*}

Signals that provide unique features that can be used for identification can also be used to boost the performance of emotion classification. When there is an absence of identifiable features, the emotion classification performance is detrimentally impacted.

\subsubsection*{Observation 6 (Comparison)}
Comparison between the different signals provides further conclusions. In particular, accelerometer data is a common signal collected by wearable devices and smartphones, while physiological signals such as ECG are not as readily available on smartphones. When we compare the performance between ACC data and other physiological signals, we implicitly compare the performance of the sensor devices. Specifically, the usage of accelerometer data achieves accuracies of 84.7\% and 97.3\% for the chest and wrist sensor, respectively, as shown in Table \ref{tab:emo}. 

\begin{eqnarray*}
	\underbrace{\texttt{\small Accuracy \textbf{ACC}}}_{\text{\it Stress state signal}}
	= \left\{
	\begin{tabular}{l}
		{\small 84.7	$\pm$	9.2,~\emph{Chest}} \\
		{\small 97.3	$\pm$	9.2,~ \emph{Wrist}} \\
	\end{tabular} 
	\right .
\end{eqnarray*}

This performance is comparable to the best performing signals, Resp and BVP. There is a greater disparity in performance between the Chest-ACC and Chest-Resp, as opposed to the Wrist-ACC and Wrist-BVP. This is most likely the result of accelerometer data being more useful near the hand as opposed to the chest which has a lower degree of movement.

%****SECTION*******SECTION*********SECTION*********SECTION****
\section{Conclusions}\label{sec:Summary}
%****SECTION*******SECTION*********SECTION*********SECTION****
This study represents the \textbf{first attempt to design the post-pandemic e-coaching approach that includes epidemiological projections}. This paper is motivated by the current COVID-19 pandemic and emphasizes the extreme demand for e-coaching. It reflects the various technology-society gaps in e-coaching. In our approach, the EMC standard conceptual platform is chosen for epidemiological projections. Four EMC pillars, i.e. Mitigation, Preparedness, Response, and Recovery, provide strong systematic requirements to e-coaching. The main benefit of this standardization is that the e-coaching can be considered as a part of computational epidemiology, -- a theoretical platform to counter epidemics and pandemics.

Secondly, the \textbf{mental state of e-coaching users in pandemic became a factor of critical importance}. In previous works on e-coaching, researchers were able only to hypothesize about psychological factors. Nowadays, evidence has accumulated confirming that the stress of the e-coached users is a powerful factor impacting the e-coaching tactic and strategy. 
	
Some intrinsic properties of e-coaching networks can be useful for counter-epidemic measures, in particular \cite{[Schmidt-2019],[Melillo-2015],[Kamphorst-2017]}, $(a)$ {scalability} that mitigates the healthcare system performance degradation,  $(b)$ {user distancing} that prevents spreading diseases, $(c)$ {continuous health monitoring} that enable real-time epidemiological landscape, and $(d)$ {personalized healthcare support} that is a common practice in pandemic. These properties are amplified by the integration of e-coaching into the smart-city concept \cite{[Celesti-2020]}.

These general properties have various perspectives for constructing a counter-epidemic framework for e-coaching systems. 
	
Our key recommendation to the designers of the e-coaching systems is to include the following \textbf{mandatory mechanisms}:
\begin{enumerate}
	\item \textbf{EMC taxonomical view}, i.e. e-coaching should be considered as a component of epidemiological surveillance and epidemiological intelligence. 
	\item \textbf{Continuous stress monitoring}, i.e. e-coaching tactic, and, consequently, strategy, must be adjusted once the user's stress state is detected. 
\end{enumerate}

There are several other findings in this paper that can be useful for researchers. In particular, we formulated the recommendations on implementation based on our experience in the area of deep learning. These recommendations are provided in Table \ref{tab:e-coaching-dln}. Note that a successful application of this learning mechanism should satisfy the specific requirements of the designed e-coaching system within a designated application.

Our study is limited to consideration of only one of the technology-society gaps, i.e. continuous stress monitoring in e-coaching. There are several other problems that are beyond this paper but are worthy of future research, in particular:
\begin{enumerate}
	\item [$-$] Intelligent agent, also called intelligent Decision Support Assistant (DSA), is a well-identified trend in e-coaching. DSA can be embodied using, in particular, an avatar technology \cite{vermetten2020using} for online coaching for post-traumatic stress disorder patients. In \cite{[Yanush-2019]}, an intelligent DSA was proposed for security applications. Another typical example of the DSA for e-coaching is a therapy prescription \cite{ruggeri-2020}. Typical DSA applications address a so-called AI bias \cite{[Whittaker-AI-now-Report-2018]}.
	\item [$-$] Risk, trust, and biases are acceptable measurements of e-coaching processes. In \cite{[Lai-2020]}, a formalization of risk, trust, and biases based on the notion of causality for intelligent automatic assessments were provided. We state that these measurements are of critical importance in e-coaching. 
\end{enumerate}
 
\section*{Acknowledgments}

\begin{small}
This Project was partially supported by Natural Sciences and Engineering Research Council of Canada (NSERC) through the grant ``Biometric intelligent interfaces''; and by the Department of National Defence's Innovation for Defence Excellence and Security (IDEaS) program, Canada.
\end{small}

\end{document}